\documentclass[12pt]{article}
\usepackage{graphicx}
\usepackage{latexsym}
\usepackage{amscd}
\usepackage[cp1251]{inputenc}
\usepackage[english,russian]{babel}
\usepackage{amsmath,  eucal}
\usepackage{amssymb}
\usepackage{indentfirst}
\usepackage[small,centerlast]{caption2}
\usepackage{longtable}
\usepackage[dvipsnames]{color}
\usepackage{cite}
\graphicspath{{figure/}}

\makeatletter
\newcommand{\const}{\mathop{\rm const\, }}

\makeatother {

\makeatletter
\renewcommand{\section}{\@startsection {section}{1}{\z@}%
                                   {-3.5ex \@plus -1ex \@minus -.2ex}%
                                   {2.3ex \@plus.2ex}%
                                   {\normalfont\Large\uppercase}}
\renewcommand{\subsection}{\@startsection{subsection}{2}{\z@}%
                                     {-3.25ex\@plus -1ex \@minus -.2ex}%
                                     {1.5ex \@plus .2ex}%
                                     {\normalfont\large\itshape}}
\renewcommand{\subsubsection}{\@startsection{subsubsection}{3}{1em}%
                                     {-3.25ex\@plus -1ex \@minus -.2ex}%
                                     {-1.5em \@plus .2em}%
                                     {\normalfont\normalsize\bfseries}}

\makeatother

\begin{document}
\renewcommand{\refname}{\begin{center}\bf REFERENCES\end{center}}
\newcommand{\mc}[1]{\mathcal{#1}}
\newcommand{\E}{\mc{E}}
\thispagestyle{empty} \large
\renewcommand{\abstractname}{ Abstract}
\centerline{\textbf{\Large  PLASMA DYNAMICS}}

 \begin{center}
{\textbf{Generation of longitudinal electric current
by the transversal electromagnetic field in collisional plasma}}
\medskip
\end{center}

\begin{center}
  \bf A. V. Latyshev\footnote{$avlatyshev@mail.ru$} and
  A. A. Yushkanov\footnote{$yushkanov@inbox.ru$}
\end{center}\medskip

\begin{center}
{\it Faculty of Physics and Mathematics,\\ Moscow State Regional
University, 105005,\\ Moscow, Radio str., 10A}
\end{center}\medskip

\begin{abstract}

Kinetic Vlasov equation for collisional plasmas
with BGK (Bhatnagar-Gross-Krook) collision integral is used.
The case of arbitrary temperature (i.e.
arbitrary degree of degeneration of electronic gas) is considered.

From kinetic Vlasov equation the distribution  function
in square-law appro\-xi\-ma\-tion on size of transversal
electromagnetic field  is received.
The formula for calculation  electric current
is deduced. This formula contains one-dimension quad\-ra\-ture.

It is shown, that the nonlinearity account leads to occurrence
the longitudinal electric current directed along the wave vector.
This longitudinal current is perpen\-di\-cu\-lar to the known
transversal classical current, received at the linear analysis.

When frequency of collisions tends to zero, all received
results for collisional  plasma pass in known corresponding
formulas for collisionless plasma.

The case of small values of wave number is considered.
It is shown, that
the received quantity of longitudinal current
at tendency of frequency of collisions to zero also passes
in known corresponding expression of  current for
collisionless plasmas.

Graphic research of dimensionless density of the current
depending on wave number, frequency of oscillations of
electromagnetic field and frequency of electron collisions with
plasma particles is carried out.
\end{abstract}

\section*{\bf Introduction}

In the present work formulas for calculation electric current
in classical collisional Fermi---Dirac plasma are deduced.

At the solution of the kinetic Vlasov equation we consider as in expansion
distribution function, and in expansion of size of the
self-consistent electro\-mag\-netic field the quan\-ti\-ties proportional
to square of intensity of external electromagnetic field.

At such nonlinear approach it appears, that an electric current
рas two nonzero components. One component of electric
current is directed along intensity of electric field.
This electric current component is precisely the same,
as well as in the linear analysis. It is "transversal"\, current.
Thus, in linear approach we receive the known
expression of transversal electric current.

The second nonzero  electric current component has the second
order of infinitesimality concerning intensity of size of
electromagnetic field.
The second electric current component is directed along the wave
vector. This current is orthogonal to the first  component.
It is "longitudinal"\,  current.

Generating in plasma of the longitudinal current by the
transversal electro\-magnetic field comes to light the nonlinear analysis
of interactions of the electromagnetic field with plasma.

Nonlinear effects in plasma  already long time
\cite{Gins} -- \cite{Shukla1} are studied.

In works \cite{Gins} and \cite{Zyt} nonlinear effects  in
plasma are studied. In work \cite{Zyt} the nonlinear current was used, in
particular, in questions of probability of decay processes. We will note,
that in work \cite{Zyt2} it is underlined existence
of nonlinear current along a wave vector
(see the formula (2.9) from \cite{Zyt2}).

Quantum plasma was studied in works \cite{Lat1} - \cite{Lat9}.
Collisional quantum plasma has started to be studied in work
\cite{Mermin}. Then quantum collisional plasma
was studied in our works \cite{Lat2} - \cite{Lat6}.
In this works quantum collisional plasma with variable collision frequency
was studied.
In works \cite{Lat7} - \cite{Lat9}
generating of  longitudinal current by transversal
electromagnetic field in classical and quantum
Fermi---Dirac plasma \cite{Lat7}, in Maxwellian plasma
\cite{Lat8} and in degenerate plasma \cite{Lat9} was investigated.


In the present work formulas for calculation electric current are deduced
in classical collisionless plasma
at any temperature  (at the any degrees of degeneration of electronic gas).

\section*{\bf 1 \quad Solution of Vlasov equation}

Let us show, that in case of the classical plasma described by the
Vlasov equation, the longitudinal current is generated and we will
calculate its density. On existence of this current was specified
more half a century ago \cite{Zyt2}.
We take Vlasov equation describing behaviour of
collisional plasmas with integral of collisions BGK
(Bhatnagar, Gross and Krook)
$$
\dfrac{\partial f}{\partial t}+\mathbf{v}\dfrac{\partial f}{\partial
\mathbf{r}}+
e\bigg(\mathbf{E}+
\dfrac{1}{c}[\mathbf{v},\mathbf{H}]\bigg)
\dfrac{\partial f}{\partial\mathbf{p}}=\nu(f^{(0)}-f).
\eqno{(1.1)}
$$

In equation (1.1) $f=f({\bf r},{\bf v},t)$ is the distribution electron
function of plasma, ${\bf E}, {\bf H}$ are the components of
electromagnetic field, $c$ is the light velocity,
${\bf p}=m{\bf v}$ is the electron momentum,
${\bf v}$ is the electron velocity, $\nu$ is the effective
electron collision frequency with plasma particles,
$f^{(0)}=f_{eq}({\bf r},v)$ (eq $\equiv$ equilibrium)
is the local equilibrium Fermi---Dirac distri\-bu\-tion,\medskip
$$
f_{eq}({\bf r},v)=\Big[1+\exp\dfrac{\E-\mu({\bf r})}{k_BT}\Big]^{-1}=
$$
$$
=\big[1+\exp(P^2-\alpha({\bf r}))\big]^{-1}=f_{eq}({\bf r},P),
$$\medskip
$\E={mv^2}/{2}$ is the electron energy, $\mu({\bf r})$ is the chemical potential
of electron gas, $k_B$ is the Boltzmann constant, $T$
is the plasma temperature, ${\bf P}={{\bf P}}/{p_T}$ is the
dimensionless electron momentum,  $p_T=mv_T$,
$v_T$ is the thermal electron velocity,
$$
v_T=\sqrt{\dfrac{2k_BT}{m}}, \qquad
\alpha({\bf r})=\dfrac{\mu({\bf r})}{(k_BT)}
$$
is the dimensionless chemical potential,
$$
k_BT=\E_T=\dfrac{mv_T^2}{2}
$$
is the thermal kinetic electron energy.

Let us consider, that in plasma there is an electromagnetic field,
repre\-sen\-ting the running harmonious wave
$$
{\bf E}={\bf E}_0e^{i({\bf kr}-\omega t)}, \qquad
{\bf H}={\bf H}_0e^{i({\bf kr}-\omega t)}.
$$

Let us consider, that the wave vector ${\bf k}$ is
orthogonal  to potential of the electromagnetic field,
$$
{\bf k A}({\bf r,t})=0.
$$

Electric and magnetic fields are connected with the vector potential
by equalities
$$
\mathbf{E}=-\dfrac{1}{c}\dfrac{\partial \mathbf{A}}{\partial t},
\;\qquad
\mathbf{H}={\rm rot} \mathbf{A}.
$$

For definiteness we will consider, that the wave vector is directed
along an axis $x $, and the electric field is directed along an axis $y $,
i.e.
$$
{\bf k}=k(1,0,0), \qquad {\bf E}=E_y(x,t)(0,1,0).
$$

Hence,
$$
\mathbf{E}=-\dfrac{1}{c}\dfrac{\partial \mathbf{A}}{\partial t}
=\dfrac{i\omega}{c}\mathbf{A},
$$
$$
{\bf H}=\dfrac{ck}{\omega}E_y\cdot(0,0,1),\qquad
{\bf [v,H}]=\dfrac{ck}{\omega}E_y\cdot (v_y,-v_x,0),
$$
$$
e\bigg(\mathbf{E}+\dfrac{1}{c}[\mathbf{v},\mathbf{H}]\bigg)
\dfrac{\partial f}{\partial\mathbf{p}}=
\dfrac{e}{\omega}E_y\Big[kv_y\dfrac{\partial f}{\partial p_x}+
(\omega-kv_x)\dfrac{\partial f}{\partial p_y}\Big],
$$
and also
$$
[\mathbf{v,H}]\dfrac{\partial f_0}{\partial \mathbf{p}}=0,\quad
\text{because}\quad
\dfrac{\partial f_0}{\partial \mathbf{p}}\sim \mathbf{v}.
$$

Let us consider linearization of locally equilibrium function
of distribution
$$
f_{eq}(P,x)=f_0(P)+g(P)\delta \alpha(x),
$$
where
$$
f_0(P)=\big[1+e^{P^2-\alpha}\big]^{-1},
$$
$$
\alpha(x)=\alpha+\delta \alpha(x),\qquad \alpha=\const,
$$
$$
g(P)=\dfrac{\partial f_0(P)}{\partial \alpha}=
\dfrac{e^{P^2-\alpha}}{(1+e^{P^2-\alpha})^2}.
$$
The equation (1.1) can be copied now in the following form
$$
\dfrac{\partial f}{\partial t}+v_x\dfrac{\partial f}{\partial x}+
\dfrac{eE_y}{\omega}\Big[kv_y\dfrac{\partial f}{\partial p_x}+
(\omega-kv_x)\dfrac{\partial f}{\partial p_y}\Big]+\nu f=
$$
$$
=\nu f_0(P)+g(P)\nu \delta \alpha(x).
\eqno{(1.2)}
$$

Size $ \delta \alpha (x) $ we will find from the law of preservation of
particles number
$$
\int (f_{eq}-f)\dfrac{2d^3p}{(2\pi \hbar)^2}=0.
$$

From this conservation law we receive that
$$
\delta \alpha(x)\int g(P)\dfrac{2d^3p}{(2\pi \hbar)^2}=\int [f-f_0(P)]
\dfrac{2d^3p}{(2\pi \hbar)^2}.
$$
From this equation we obtain that
$$
\delta \alpha(x)=\dfrac{\displaystyle\int [f-f_0(P)] d^3P}
{\displaystyle\int g(P)d^3P}.
$$

We notice that
$$
\int g(P)d^3P=2\pi \int\limits_{0}^{\infty}\dfrac{dP}{1+e^{P^2-\alpha}}=
\pi \int\limits_{-\infty}^{\infty}\dfrac{dP}{1+e^{P^2-\alpha}}=
\pi \hat f_0(\alpha),
$$
where
$$
\hat f_0(\alpha)= \int\limits_{-\infty}^{\infty}\dfrac{dP}{1+e^{P^2-\alpha}}
=2\int\limits_{0}^{\infty}\dfrac{dP}{1+e^{P^2-\alpha}}.
$$

Therefore
$$
\delta \alpha(x)=\dfrac{1}{\pi\hat f_0(\alpha)}\int [f-f_0(P)] d^3P.
$$

The equation (1.2) will be transformed now to the integrated equation
$$
\dfrac{\partial f}{\partial t}+v_x\dfrac{\partial f}{\partial x}+
\nu f=\nu f_0(P)-\dfrac{eE_y}{\omega}\Big[kv_y\dfrac{\partial f}{\partial p_x}+
(\omega-kv_x)\dfrac{\partial f}{\partial p_y}\Big]+
$$
$$
+\nu g(P)\dfrac{1}{\pi\hat f_0(\alpha)}\int [f-f_0(P)] d^3P.
\eqno{(1.3)}
$$

Let us search for the solution of the equation (1.3) in the form
$$
f=f_0(P)+f_1+f_2,
\eqno{(1.4)}
$$
where
$$
f_1\sim E_y\sim e^{i(kx-\omega t)},
$$
$$
f_2\sim E_y^2\sim e^{2i(kx-\omega t)}.
$$

Let us operate with  method consecutive approximations, considering
as small parameter size of intensity of electric field.
Then the equation (1.3)  with help (1.4) equivalent to  the
following equations
$$
\dfrac{\partial f_1}{\partial t}+
v_x\dfrac{\partial f_1}{\partial x}+\nu f_1=
$$
$$
=-\dfrac{eE_y}{\omega}\Bigg[kv_y\dfrac{\partial f_0}{\partial p_x}+
(\omega-kv_x)\dfrac{\partial f_0}{\partial p_y}\Bigg]
+\nu g(P)\dfrac{1}{\pi \hat f_0(\alpha)}\int f_1 d^3P.
\eqno{(1.5)}
$$ \bigskip
and
$$
\dfrac{\partial f_2}{\partial t}+
v_x\dfrac{\partial f_2}{\partial x}+\nu f_2=
$$
$$
=-\dfrac{eE_y}{\omega}\Bigg[kv_y\dfrac{\partial f_1}{\partial p_x}+
(\omega-kv_x)\dfrac{\partial f_1}{\partial p_y}\Bigg]
+\nu g(P)\dfrac{1}{\pi \hat f_0(\alpha)}\int f_2 d^3P.
\eqno{(1.6)}
$$ \bigskip

From equation (1.5) we obtain that
$$
(\nu-i\omega+ikv_x)f_1=
$$
$$
=-\dfrac{eE_y}{\omega}
\Bigg[kv_y\dfrac{\partial f_0}{\partial p_x}+
(\omega-kv_x)\dfrac{\partial f_0}{\partial p_y}\Bigg]
+\nu g(P)A_1.
$$

Here
$$
A_1=\dfrac{1}{\pi \hat f_0(\alpha)}\int f_1 d^3P.
\eqno{(1.7)}
$$

Let us enter dimensionless parametres
$$
\Omega=\dfrac{\omega}{k_Tv_T},\qquad y=\dfrac{\nu}{k_Tv_T},
\qquad q=\dfrac{k}{k_T}.
$$

Here $q $ is the dimensionless wave number,
$k_T =\dfrac {mv_T} {\hbar} $ is the thermal wave number, $ \Omega $
is the dimensionless frequency of oscillations of the electro\-mag\-ne\-tic field.

In the previous equation we will pass to dimensionless parametres
$$
i(qP_x-z)f_1=$$$$=-\dfrac{eE_y}{\Omega k_Tp_Tv_T}
\Bigg[qP_y\dfrac{\partial f_0}{\partial P_x}+
(\Omega-qP_x)\dfrac{\partial f_0}{\partial P_y}\Bigg]
+y g(P)A_1.
\eqno{(1.8)}
$$

Here
$$
z=\Omega+iy=\dfrac{\omega+iy}{k_Tv_T}.
$$

We notice that
$$
\dfrac{\partial f_0}{\partial P_x}\sim P_x,\qquad
\dfrac{\partial f_0}{\partial P_y}\sim P_y.
$$

Hence
$$
\Bigg[qP_y\dfrac{\partial f_0}{\partial P_x}+
(\Omega-qP_x)\dfrac{\partial f_0}{\partial P_y}\Bigg]=
\Omega\dfrac{\partial f_0}{\partial P_y}.
$$

Now from the equation (1.8) we find, that
$$
f_1=\dfrac{ieE_y}{k_Tp_Tv_T}\cdot\dfrac{\partial f_0/\partial P_y}
{qP_x-z}-iy\cdot\dfrac{g(P)}{qP_x-z}A_1.
\eqno{(1.9)}
$$

Let us substitute (1.9) in the equation (1.7). We receive equality
$$
A_1\Bigg(1+\dfrac{iy}{2\pi \hat f_0(\alpha)}\int \dfrac{g(P)d^3P}
{qP_x-z}\Bigg)=\dfrac{ieE_y}{k_Tp_Tv_T}\int
\dfrac{\partial f_0/\partial P_y}{qP_x-z}d^3P.
$$

It is easy to see, that integral in the right part of this equality
is equal to zero. Hence
$$
A_1=0.
$$
Thus, it agree
(1.9) function $f_1$ is constructed and defined by equality
$$
f_1=\dfrac{ieE_y}{k_Tp_Tv_T}\cdot\dfrac{\partial f_0/\partial P_y}
{qP_x-z}.
\eqno{(1.10)}
$$

In the second approximation we  substitute $f_1 $  according (1.10)
in the equation (1.6).

We will receive the equation
$$
(\nu-2i\omega+2ikv_x)f_2=$$$$
-\dfrac{ie^2E_y^2}{k_Tp_Tv_T\omega}\Big[kv_y\dfrac{\partial}{\partial p_x}
\Big(\dfrac{\partial f_0/\partial P_y}{qP_x-z}\Big)+
(\omega-kv_x)\dfrac{\partial }
{\partial p_y}\Big(\dfrac{\partial f_0/\partial
P_y}{qP_x-z}\Big)\Big]+
$$
$$
+\nu g(P)A_2.
$$

Here
$$
A_2=\dfrac{1}{\pi\hat f_0(\alpha)}\int f_2d^3P.
\eqno{(1.11)}
$$

Let us pass in this equation to dimensionless parameters. We
receive the equation
$$
2i(qP_x-x-\dfrac{iy}{2})f_2=$$$$=-\dfrac{ie^2E_y^2}{\Omega k_T^2p_T^2v_T^2}
\Big[qP_x\dfrac{\partial}{\partial P_x}
\Big(\dfrac{\partial f_0/\partial P_y}{qP_x-z}\Big)+
(\Omega-qP_x)\dfrac{\partial }
{\partial P_y}\Big(\dfrac{\partial f_0/\partial
P_y}{qP_x-z}\Big)\Big]+
$$
$$
+yg(P)A_2.
$$
Let us designate
$$
z'=\Omega+\dfrac{iy}{2}=\dfrac{\omega}{k_Tv_T}+i\dfrac{\nu}{2k_Tv_T}=
\dfrac{\omega+i \nu/2}{k_Tv_T}.
$$

From last equation it is found
$$
f_2=-\dfrac{e^2E_y^2}{2k_T^2p_T^2v_T^2 \Omega}
\Bigg[qP_y\dfrac{\partial}{\partial P_x}
\Big(\dfrac{\partial f_0/\partial P_y}{qP_x-z}\Big)+
\dfrac{\Omega-qP_x}{qP_x-z}\dfrac{\partial^2f_0}{\partial P_y^2}\Bigg]
\dfrac{1}{qP_x-z'}-
$$
$$
-\dfrac{iy}{2}\cdot\dfrac{g(P)}{qP_x-z'}A_2.
\eqno{(1.12)}
$$

For finding $A_2$ we will substitute (1.12) in (1.11). From the received
relation it is found $A_2$
$$
A_2=-\dfrac{e^2E_y^2}{2k_T^2p_T^2v_T^2\Omega}\cdot\dfrac{J_1}
{\pi \hat f_0(\alpha)+\dfrac{iy}{2}J_0}.
$$

Here
$$
J_0=\int\dfrac{g(P)d^3P}{qP_x-z'},
$$
$$
J_1=\int\Bigg[qP_y\dfrac{\partial}{\partial P_x}
\Big(\dfrac{\partial f_0/\partial P_y}{qP_x-z}\Big)+
\dfrac{\Omega-qP_x}{qP_x-z}\dfrac{\partial^2f_0}{\partial P_y^2}\Bigg]
\dfrac{d^3P}{qP_x-z'}
$$

Substituting the found value $A_2$ in (1.12), definitively
we find function $f_2$ in the explicit form
$$
f_2=-\dfrac{e^2E_y^2}{2k_T^2p_T^2v_T^2 \Omega}
\Bigg[qP_y\dfrac{\partial}{\partial P_x}
\Big(\dfrac{\partial f_0/\partial P_y}{qP_x-z}\Big)+
\dfrac{\Omega-qP_x}{qP_x-z}\dfrac{\partial^2f_0}{\partial P_y^2}\Bigg]
\dfrac{1}{qP_x-z'}+
$$
$$
+\gamma\dfrac{e^2E_y^2}{2k_T^2p_T^2v_T^2 \Omega}
\cdot\dfrac{g(P)}{qP_x-z'},
\eqno{(1.13)}
$$
where
$$
\gamma=\dfrac{(iy/2)J_1}{\pi\hat f_0(\alpha)+(iy/2)J_0}.
\eqno{(1.14)}
$$

\section*{\bf 2\quad Density of electric current}

Let us find electric current density
$$
\mathbf{j}=e\int \mathbf{v}f \dfrac{2d^3p}{(2\pi\hbar)^3}.
\eqno{(2.1)}
$$

From equalities (1.4) -- (1.6) it is visible, that the vector
of current density has two nonzero components
$$
\mathbf{j}=(j_x,j_y,0).
$$

Here $j_y$ is the density of transversal current,
$$
j_y=e\int v_yf \dfrac{2d^3p}{(2\pi\hbar)^3}=
e\int v_yf_1 \dfrac{2d^3p}{(2\pi\hbar)^3}.
$$

This current is directed along an electric field, its density
it is defined only by the first approximation of distribution
function.

The second approximation of  distribution function
the contribution to current density does not bring.

The density of transversal current is defined by equality
$$
j_y=\dfrac{2ie^2p_T^2}{(2\pi\hbar)^3k_T}E_y(x,t)
\int\dfrac{(\partial f_0/\partial P_y)P_y}{qP_x-z}d^3P.
$$

This current is proportional to the first degree of size of
electric field intensity.

For density of longitudinal current according to its definition it is had
$$
j_x=e\int v_xf\dfrac{2d^3p}{(2\pi\hbar)^3}=
e\int v_xf_2\dfrac{2d^3p}{(2\pi\hbar)^3}=
\dfrac{2ev_Tp_T^3}{(2\pi\hbar)^3}\int P_xf_2d^3P.
$$

By means of (1.6) from here it is received, that
$$
j_x=\dfrac{e^3E_y^2m}{(2\pi\hbar)^3k_T^2\Omega}\Bigg[-\int
\Bigg[qP_y\dfrac{\partial}{\partial P_x}
\Big(\dfrac{\partial f_0/\partial P_y}{qP_x-z}\Big)+
\dfrac{x-qP_x}{qP_x-z}\dfrac{\partial^2f_0}{\partial P_y^2}\Bigg]
\dfrac{P_xd^3P}{qP_x-z'}+
$$
$$
+\gamma\int\dfrac{P_xg(P)d^3P}{qP_x-z'}
\Bigg].
\eqno{(2.2)}
$$

In integral from the second composed from square bracets (2.2) internal
integral on $P_y $ it is equal to zero:
$$
\int\limits_{-\infty}^{\infty}\dfrac{\partial^2f_0}{\partial P_y^2}dP_y
=\dfrac{\partial f_0}{\partial P_y}\Bigg|_{P_y=-\infty}^{P_y=+\infty}=0.
$$

In the first integral from square bracets (2.2) internal integral on $P_x $
is calculated in parts
$$
\int\limits_{-\infty}^{\infty}\dfrac{\partial}{\partial P_x}
\Big(\dfrac{\partial f_0/\partial P_y}{qP_x-z}\Big)
\dfrac{P_xdP_x}{qP_x-z'}=z' \int\limits_{-\infty}^{\infty}
\dfrac{\partial f_0/\partial P_y}{(qP_x-z')^2(qP_x-z)}dP_x.
$$

Hence, equality (2.2) becomes simpler
$$
j_x=\dfrac{e^3E_y^2m}{(2\pi\hbar)^3k_T^2\Omega}\Bigg[-z'q\int
\dfrac{P_y(\partial f_0/\partial P_y)d^3P}{(qP_x-z')^2(qP_x-z)}+
$$
$$
+\gamma \int\dfrac{P_xg(P)d^3P}{qP_x-z'}\Bigg].
$$

Internal integral on variable $P_y $ we will integrate on parts
$$
\int\limits_{-\infty}^{\infty}P_y\dfrac{\partial f_0}{\partial P_y}dP_y=
P_yf_0\Bigg|_{P_y=-\infty}^{P_y=+\infty}-
\int\limits_{-\infty}^{\infty}f_0(P)dP_y=
-\int\limits_{-\infty}^{\infty}f_0(P)dP_y.
$$

Thus, expression for the longitudinal current becomes
$$
j_x=\dfrac{e^3E_y^2m}{(2\pi\hbar)^3k_T^2\Omega}\Bigg[z'q\int
\dfrac{f_0(P)d^3P}{(qP_x-z')^2(qP_x-z)}+
$$
$$
+\gamma \int\dfrac{P_xg(P)d^3P}{qP_x-z'}\Bigg].
\eqno{(2.3)}
$$

Internal integral in plane $ (P_y, P_z) $ we will calculate in the polar
сoordi\-na\-tes
$$
\int\dfrac{f_0(P)d^3P}{(qP_x-z')^2(qP_x-z)}=\int\limits_{-\infty}^{\infty}
\dfrac{dP_x}{(qP_x-z')^2(qP_x-z)}
\int\limits_{-\infty}^{\infty}\int\limits_{-\infty}^{\infty}
f_0(P)dP_ydP_z=
$$
$$
=\pi\int\limits_{-\infty}^{\infty}
\dfrac{\ln(1+e^{\alpha-P_x^2})dP_x}{(qP_x-z')^2(qP_x-z)},
$$
since
$$
\int\limits_{-\infty}^{\infty}\int\limits_{-\infty}^{\infty}
f_0(P)dP_ydP_z=\pi\ln(1+e^{\alpha-P_x^2}).
$$

In addition
$$
\int\dfrac{P_xg(P)d^3P}{qP_x-z'}=\pi\int\limits_{-\infty}^{\infty}
\dfrac{P_xf_0(P_x)dP_x}{qP_x-z'}=
$$
$$
=\pi\int\limits_{-\infty}^{\infty}\dfrac{\tau d \tau}{(1+e^{\tau^2-\alpha})
(q\tau-z')}=\pi\int\limits_{-\infty}^{\infty}
\dfrac{e^{\alpha-\tau^2}\tau d \tau}{(1+e^{\alpha-\tau^2})
(q\tau-z')}=
$$
$$
=-\dfrac{\pi q}{2}\int\limits_{-\infty}^{\infty}
\dfrac{\ln(1+e^{\alpha-\tau^2})d\tau}{(q\tau-z')^2}.
$$

Equality (2.3) is reduced to one-dimensional integral
$$
j_x=\dfrac{\pi e^3E_y^2mq}{(2\pi\hbar)^3k_T^2\Omega}\Bigg[z'
\int\limits_{-\infty}^{\infty}
\dfrac{\ln(1+e^{\alpha-\tau^2})d\tau}{(q\tau-z')^2(q\tau-z)}-
\dfrac{\gamma}{2}\int\limits_{-\infty}^{\infty}
\dfrac{\ln(1+e^{\alpha-\tau^2})d\tau}{(q\tau-z')^2}\Bigg].
$$

Let us rewrite the previous equality in the form
$$
j_x=\dfrac{\pi e^3E_y^2mq}{(2\pi\hbar)^3k_T^2\Omega}\Big[z'J_{12}-
\dfrac{\gamma}{2}J_{02}\Big].
\eqno{(2.4)}
$$
where
$$
J_{12}=\int\limits_{-\infty}^{\infty}
\dfrac{\ln(1+e^{\alpha-\tau^2})d\tau}{(qP_x-z)(q\tau-z')^2},
$$
and
$$
J_{02}=\int\limits_{-\infty}^{\infty}
\dfrac{\ln(1+e^{\alpha-\tau^2})d\tau}{(q\tau-z')^2}.
$$

Let us return to consideration of size  $ \gamma $.
We will calculate integrals,
entering in (1.14). We will calculate the first integral
$$
J_1=\int
\Bigg[qP_y\dfrac{\partial}{\partial P_x}
\Big(\dfrac{\partial f_0/\partial P_y}{qP_x-z}\Big)+
\dfrac{x-qP_x}{qP_x-z}\dfrac{\partial^2f_0}{\partial P_y^2}\Bigg]
\dfrac{d^3P}{qP_x-z'}.
$$

As it was already specified, the integral from the second composed
is equal to zero. The second integral as well as earlier we will
calculate in parts. As a result цe receive
$$
J_1=q\int P_y \dfrac{\partial}{\partial P_x}
\Big(\dfrac{\partial f_0/\partial P_y}{qP_x-z}\Big)\dfrac{d^3P}{qP_x-z'}=
$$
$$
=q^2\int \dfrac{P_y[\partial f_0/\partial P_y]d^3P}{(qP_x-z)(qP_x-z')^2}.
$$

Now we will calculate in parts internal integral on the variable
$P_y $. As the result we receive
$$
J_1=-q^2\int\dfrac{ f_0(P)d^3P}{(qP_x-z)(qP_x-z')^2}.
$$

This integral has been calculated earlier. Hence
$$
J_1=-\pi q^2\int\limits_{-\infty}^{\infty}
\dfrac{\ln(1+e^{\alpha-\tau^2})d\tau}{(qP_x-z)(q\tau-z')^2}=
-\pi q^2 J_{12}.
$$

Let us calculate the second integral from (1.14). We have
$$
J_0=\int\dfrac{g(P)d^3P}{qP_x-z'}=\int\limits_{-\infty}^{\infty}
\dfrac{dP_x}{qP_x-z'}\int\limits_{-\infty}^{\infty}
\int\limits_{-\infty}^{\infty} g(P)dP_ydP_z.
$$

The internal double integral is equal
$$
\int\limits_{-\infty}^{\infty}
\int\limits_{-\infty}^{\infty} g(P)dP_ydP_z=\pi\dfrac{1}{1+e^{P^2-\alpha}}
=\pi f_0(P).
$$

Hence, the integral $J_0$ is equal
$$
J_0=\pi \int\limits_{-\infty}^{\infty}\dfrac{f_0(\tau)d\tau}{q\tau-z'}.
$$

So
$$
\pi\hat f_0(\alpha)+\dfrac{iy}{2}J_0=\pi
\int\limits_{-\infty}^{\infty}\dfrac{q\tau-\Omega}{q\tau-z'}
f_0(\tau)d\tau.
$$

Thus, the constant $ \gamma $ is found
$$
\gamma=-\dfrac{iy}{2}q^2 \dfrac{J_{12}}{J_{01}},
$$
where
$$
J_{01}=\int\limits_{-\infty}^{\infty}\dfrac{q\tau-\Omega}{q\tau-z'}
f_0(\tau)d\tau.
$$

Now (2.4) it is possible to present the formula in the form
$$
j_x=\dfrac{\pi e^3E_y^2mq}{(2\pi\hbar)^3k_T^2\Omega}\Big[
\Omega+\dfrac{iy}{2}+\dfrac{iy}{4}q^2\dfrac{J_{02}}{J_{01}}\Big]J_{12}.
\eqno{(2.4')}
$$

Let us find numerical density (concentration) of plasma
particles, cor\-res\-pon\-ding to Fermi---Dirac distribution
$$
N=\int f_0(P)\dfrac{2d^3p}{(2\pi\hbar)^3}=
\dfrac{8\pi p_T^3}{(2\pi\hbar)^3}\int\limits_{0}^{\infty}
\dfrac{e^{\alpha-P^2}P^2dP}{1+e^{\alpha-P^2}}=
\dfrac{k_T^3}{2\pi^2}l_0(\alpha),
$$
where $k_T$ is the thermal wave number, $k_T=\dfrac{mv_T}{\hbar}$,
$$
l_0(\alpha)=\int\limits_{0}^{\infty}\ln(1+e^{\alpha-\tau^2})d\tau.
$$

In expression before integral from (2.4) we will allocate the plasma
(Lang\-muir) frequency
$$
\omega_p=\sqrt{\dfrac{4\pi e^2N}{m}}
$$
and number density (concentration) $N$,
and last we will express through thermal wave number. We will receive
$$
{j_x}^{\rm long}=\Big(\dfrac{e\Omega_p^2}{k_Tp_T}\Big)
\dfrac{k{E_y^2}}{16\pi l_0(\alpha)\Omega}\Big[
\Omega+\dfrac{iy}{2}+\dfrac{iy}{4}q^2\dfrac{J_{02}}{J_{01}}\Big]J_{12},
\eqno{(2.5)}
$$
where
$$
\Omega_p=\dfrac{\omega_p}{k_Tv_T}=\dfrac{\hbar\omega_p}{mv_T^2}
$$
is the dimensionless plasma frequency.

Equality (2.5) we rewrite in the form
$$
j_x^{\rm long}=J(\Omega,y,q)\sigma_{l,tr}kE_y^2,
\eqno{(2.6)}
$$
where $\sigma_{l,tr}$ is the longitudinal--transversal conductivity,
$J(\Omega,y,q)$ is the dimen\-sionless part of current,
$$
\sigma_{l,tr}=
\dfrac{e \Omega_p^2}{p_Tk_T}=\dfrac{e\hbar}{p_T^2}
\Big(\dfrac{\hbar \omega_p}{mv_T^2}\Big)^2=
\dfrac{e}{k_Tp_T}\Big(\dfrac{\omega_p}{k_Tv_T}\Big)^2,
$$
$$
J(\Omega,y,q)=\dfrac{1}{16\pi l_0(\alpha)\Omega}
\Big[
\Omega+\dfrac{iy}{2}+\dfrac{iy}{4}q^2\dfrac{J_{02}}{J_{01}}\Big]J_{12}.
$$

If to enter transversal field
$$
\mathbf{E}_{\rm tr}=\mathbf{E}-\dfrac{\mathbf{k(Ek)}}{k^2}=
\mathbf{E}-\dfrac{\mathbf{q(Eq)}}{q^2},\qquad
{\bf kE}_{tr}=\dfrac{\omega}{c}[{\bf E,H}],
$$
then the equality (2.6) we will to write down in
invariant form
$$
\mathbf{j}^{\rm long}=J(\Omega,y,q)\sigma_{l,tr}{\bf k}{\bf E}_{tr}^2
=J(\Omega,y,q)\sigma_{l,tr}\dfrac{\omega}{c}[{\bf E,H}].
$$

{\sc Remark.}
From the formula (2.5) (or from (2.6)) it is visible, that at $y=0$ (or
$ \nu=0$) i.e. when collisional plasma passes in
collisionless ($z\to \Omega, z '\to \Omega $),
this formula in accuracy passes in the corresponding formula
from our work \cite{Lat7} for collisionless plasmas
$$
{j_x}^{\rm long}=\sigma_{\rm l,tr}k{E_y^2}
\dfrac{1}{16\pi l_0(\alpha)}\int\limits_{-\infty}^{\infty}
\dfrac{\ln(1+e^{\alpha-\tau^2})d\tau}{(q\tau-\Omega)^2(q\tau-\Omega)},
$$

Let us pass to consideration of the case of small values of wave num\-ber.
From expression (2.5) at small values of wave number it is received
$$
{j_x}^{\rm long}=-\sigma_{\rm l,tr}k{E_y^2}
\dfrac{1}{16\pi l_0(\alpha)\Omega zz'}
\int\limits_{-\infty}^{\infty}
\ln(1+e^{\alpha-\tau^2})d\tau=
$$
$$
=-\dfrac{\sigma_{\rm l,tr}k{E_y^2}}{8\pi \Omega zz'}=
-\dfrac{e}{8\pi m\omega}
\Big(\dfrac{\omega_p}{\omega}\Big)^2\dfrac{k{E_y^2}}
{\Big(1-i\dfrac{\nu}{\omega}\Big)\Big(1-i\dfrac{\nu}{2\omega}\Big)}.
$$

{\sc Remark.} At $\nu=0$ from this formula in accuracy turns out
the corres\-pon\-ding formula from \cite{Lat7}  for
longitudinal current in the case of small values of wave number in
collisionless plasma.

\section*{\bf 3\quad Conclusion}

On Figs. 1 and 2 we will present behaviour of real (Fig. 1)
and imaginary (Fig. 2) parts of density of the dimensionless
longitudinal current at $ \Omega=1, y=0.01$ depending on
dimensionless wave number $q $ at various values of dimensionless
chemical potential.
At small and great values of parametre $q $ curves {\it 1,2} and {\it 3}
approach and become indiscernible.
The real part has at first the minimum, and then the maximum.
With growth of the dimensionless
chemical potential the imaginary part of density of a current has one
maximum.

Further graphic research of size of density of the longitudinal current
let us spend for the case of zero chemical potential:
$ \alpha=0$ (Figs. 3 -- 6).

On Figs. 3 and 4 we will present behaviour of real (Fig. 3) and imaginary
(Fig. 4) parts of density of the longitudinal current depending on
dimen\-si\-on\-less wave numbers $q $ in the case $ \Omega=1$
at various values of dimensionless frequency of electron collisions.
At small and at great values of dimen\-sion\-less wave number
curves {\it 1,2} and {\it 3} approach and become indiscernible.

On Figs. 5 and 6 we will present behaviour of real (Fig. 5) and
imaginary (Fig. 6) parts of density of the longitudinal current
in dependence from dimensionless frequency of oscillations
of the electromagnetic field $ \Omega $
in the case $q=0.3$. At increase of dimensionless wave number
$q $ curves {\it 1,2} and {\it 3} approach and at $q> 0.7$
practically coincide.

In the present work influence of nonlinear character of
interactions of the electromagnetic field with the classical
collisional plasma is considered.

It has appeared, that presence of nonlinearity of the electromagnetic field
leads to generating of the electric current, orthogonal to
the field direction.

Further authors purpose to consider a problem of the plasma
oscillations and a problem about skin-effect with use square
vector potential in expan\-sion of  distribution function.

\clearpage

\begin{figure}[ht]\center
\includegraphics[width=16.0cm, height=9cm]{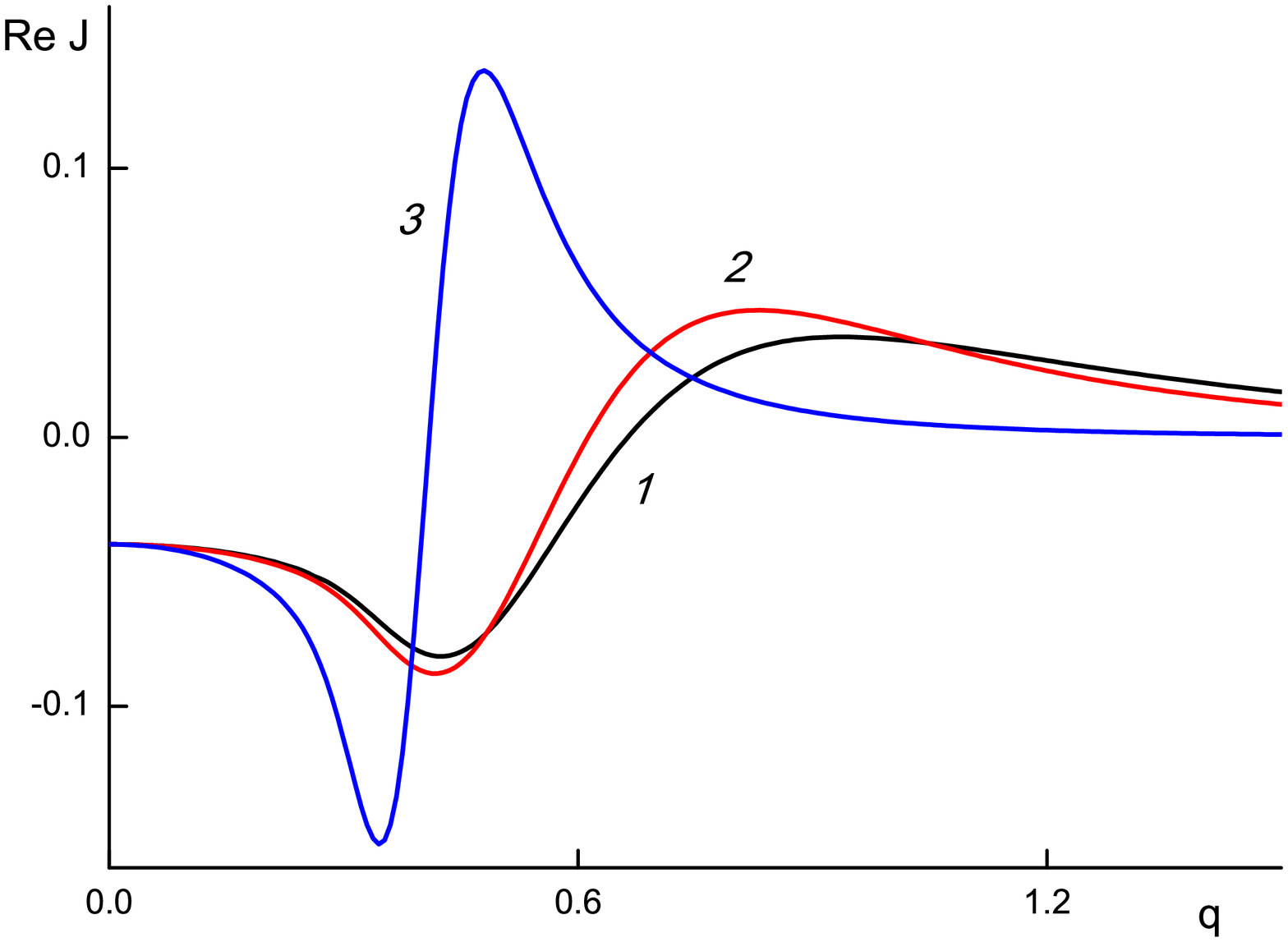}
\center{Fig. 1. Real part of dimensionless density of longitudinal current,
$\Omega=1, y=0.01$. Curves $1,2,3$ correspond to values of
dimensionless chemical potential $\alpha=-5, 0, +5$.}
\includegraphics[width=17.0cm, height=9cm]{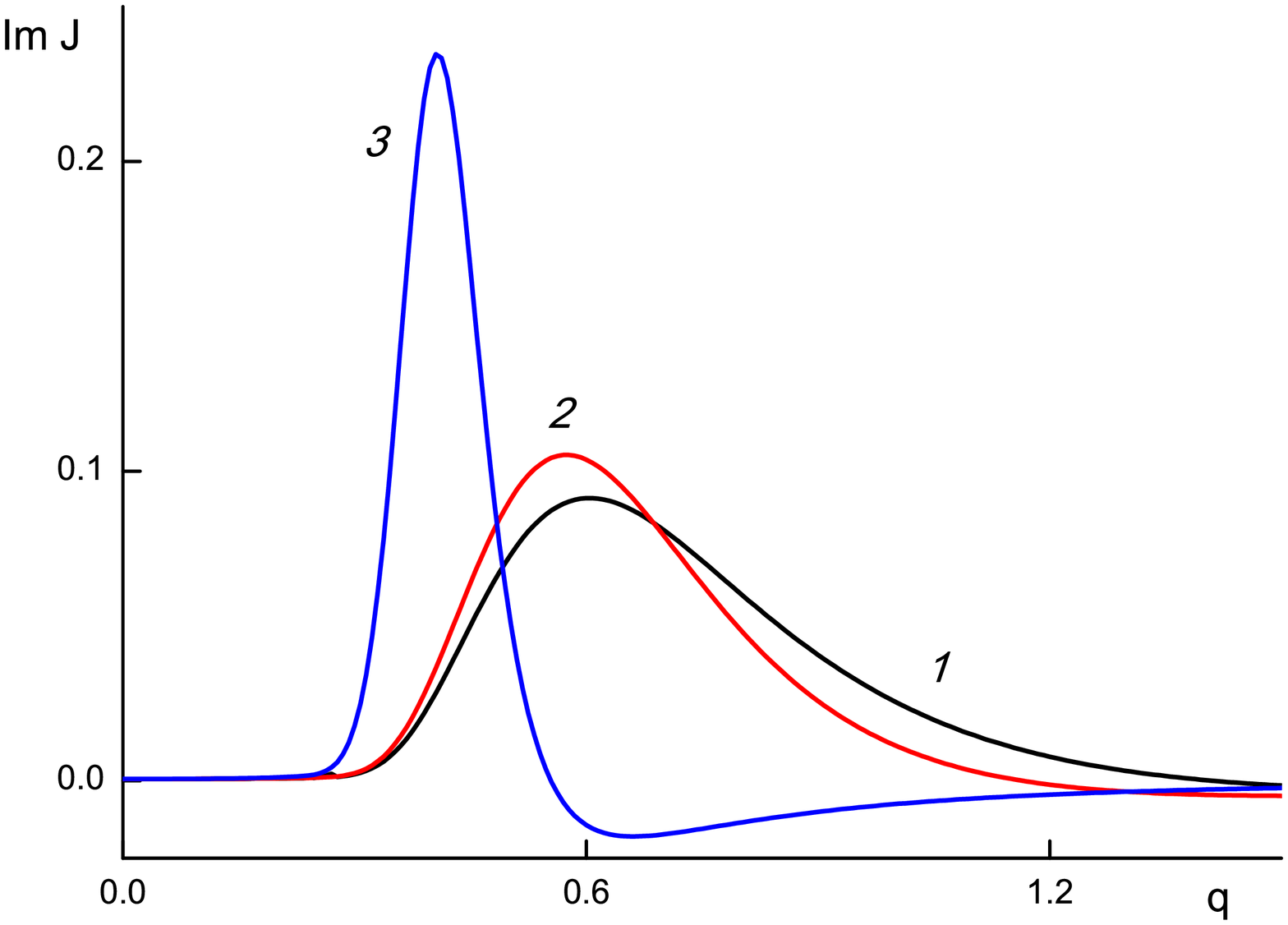}
\center{Fig. 2. Imaginary part of dimensionless density of longitudinal current,
$\Omega=1, y=0.01$. Curves $1,2,3$ correspond to values of
dimensionless chemical potential $\alpha=-5, 0, +5$.}
\end{figure}

\begin{figure}[ht]\center
\includegraphics[width=16.0cm, height=9cm]{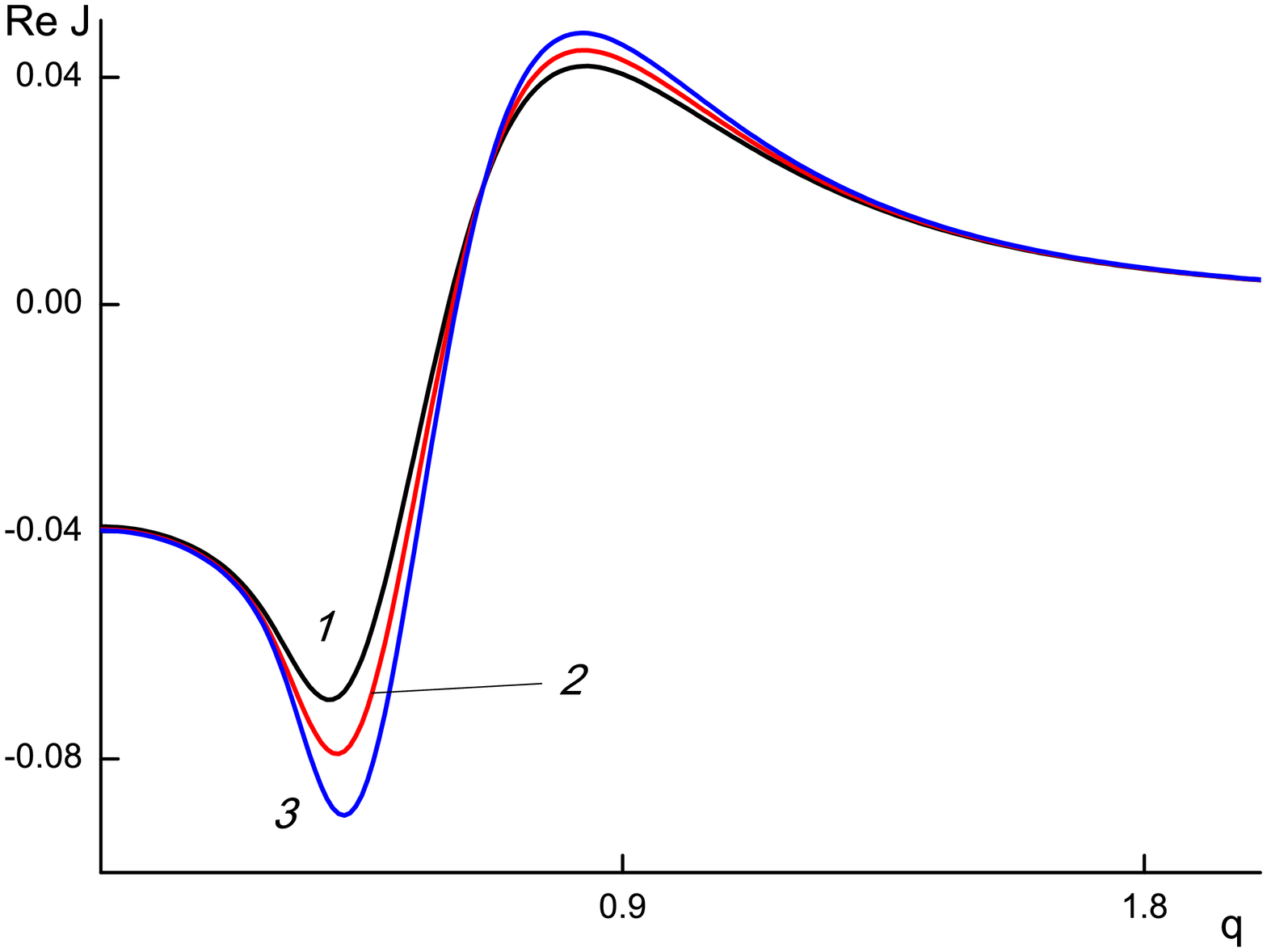}
\center{Fig. 3. Real part of dimensionless density of longitudinal current,
$\Omega=1, \alpha=0$. Curves $1,2,3$ correspond to values of
dimensionless collision frequency $y=0.001, 0.05, 0.1$.}
\includegraphics[width=17.0cm, height=9cm]{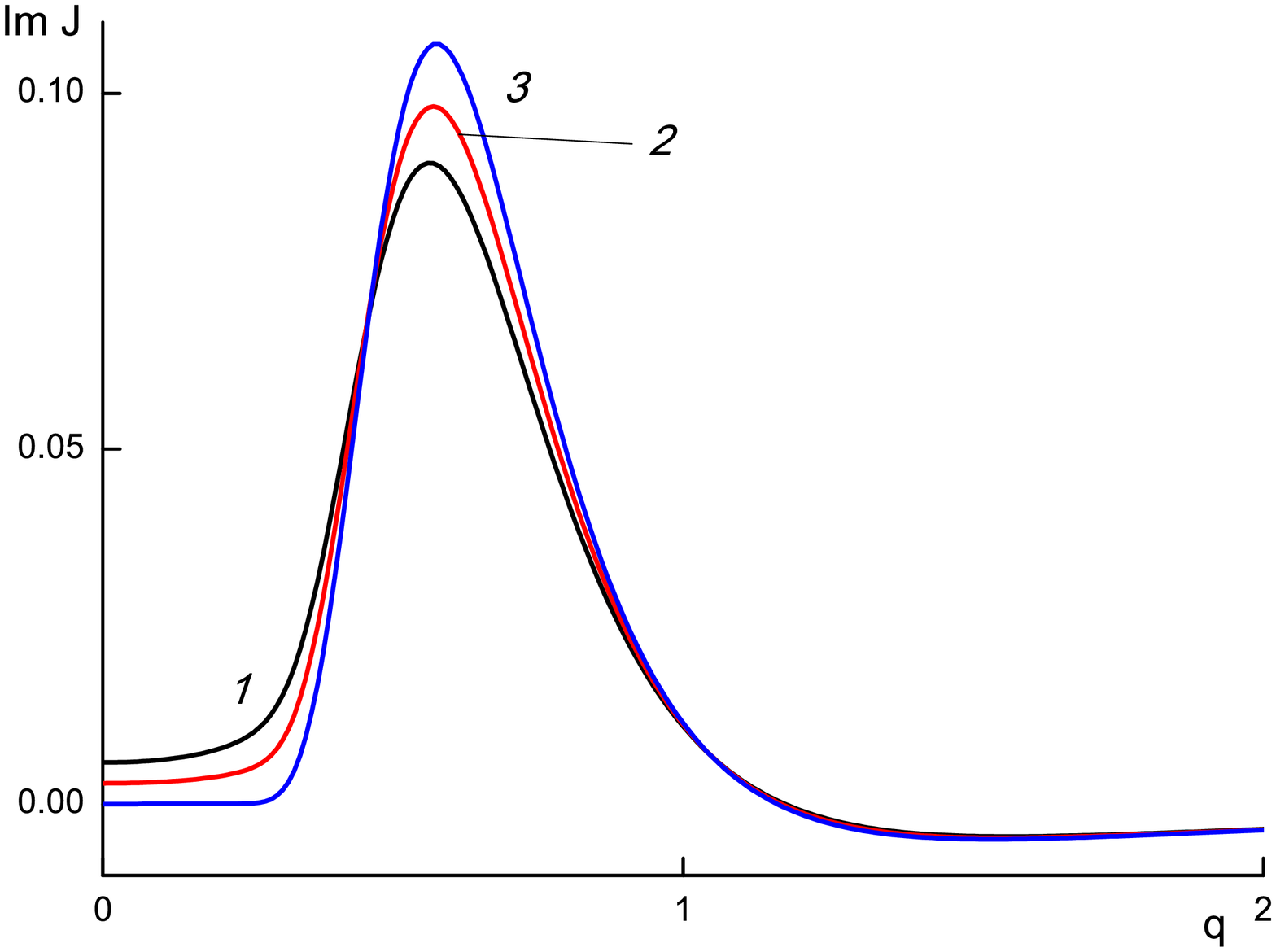}
\center{Fig. 4. Imaginary part of dimensionless density of longitudinal current,
$\Omega=1, \alpha=0$. Curves $1,2,3$ correspond to values of
dimensionless collision frequency $y=0.001, 0.05, 0.1$.}
\end{figure}

\begin{figure}[ht]\center
\includegraphics[width=16.0cm, height=9cm]{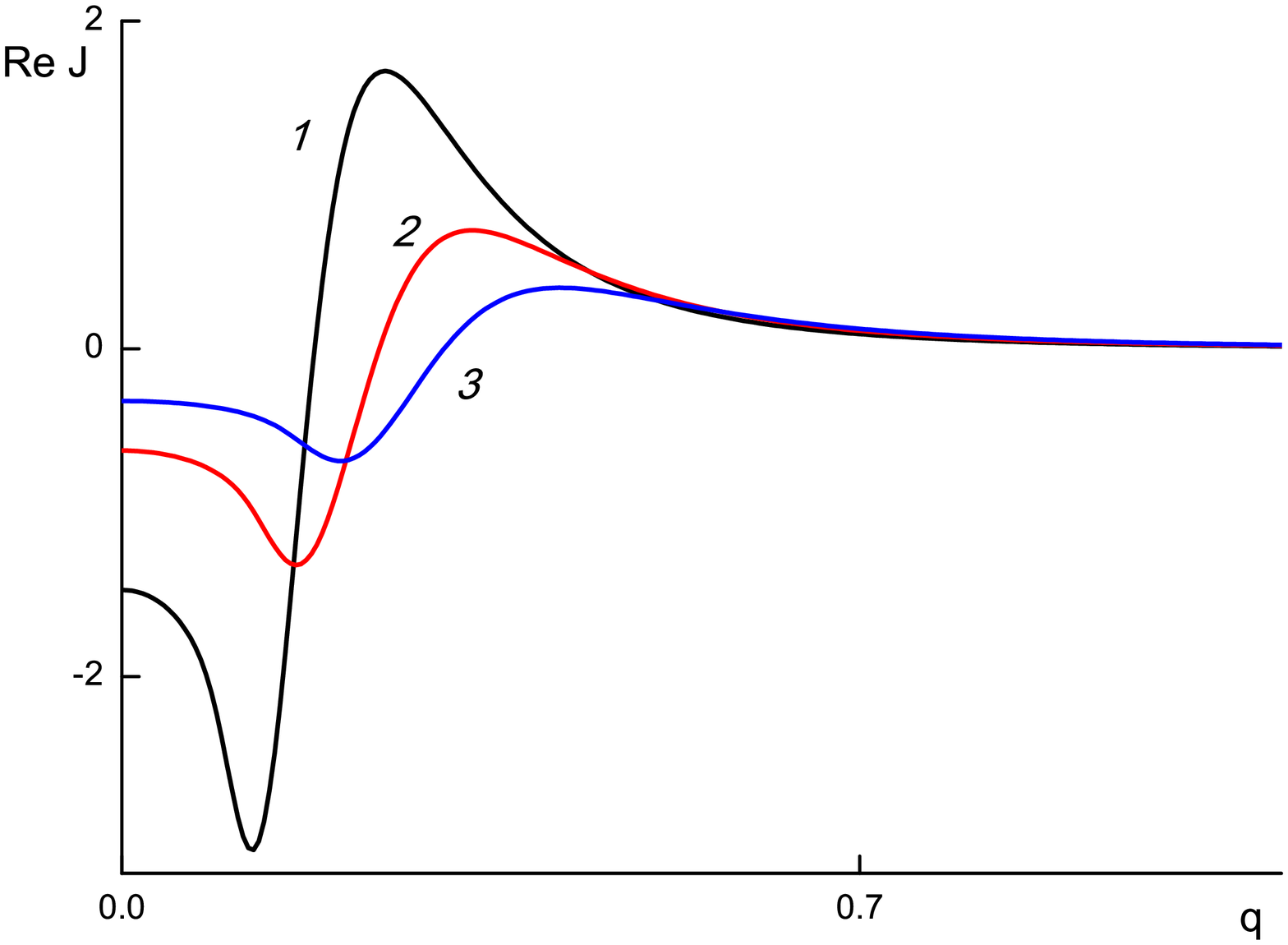}
{Fig. 5. Real part of dimensionless density of longitudinal current,
$y=0.01, \alpha=0$. Curves $1,2,3$  correspond to values of
dimensionless frequency of oscillations \\of electromagnetic field
$\Omega=0.3, 0.4, 0.5$.}
\includegraphics[width=17.0cm, height=9cm]{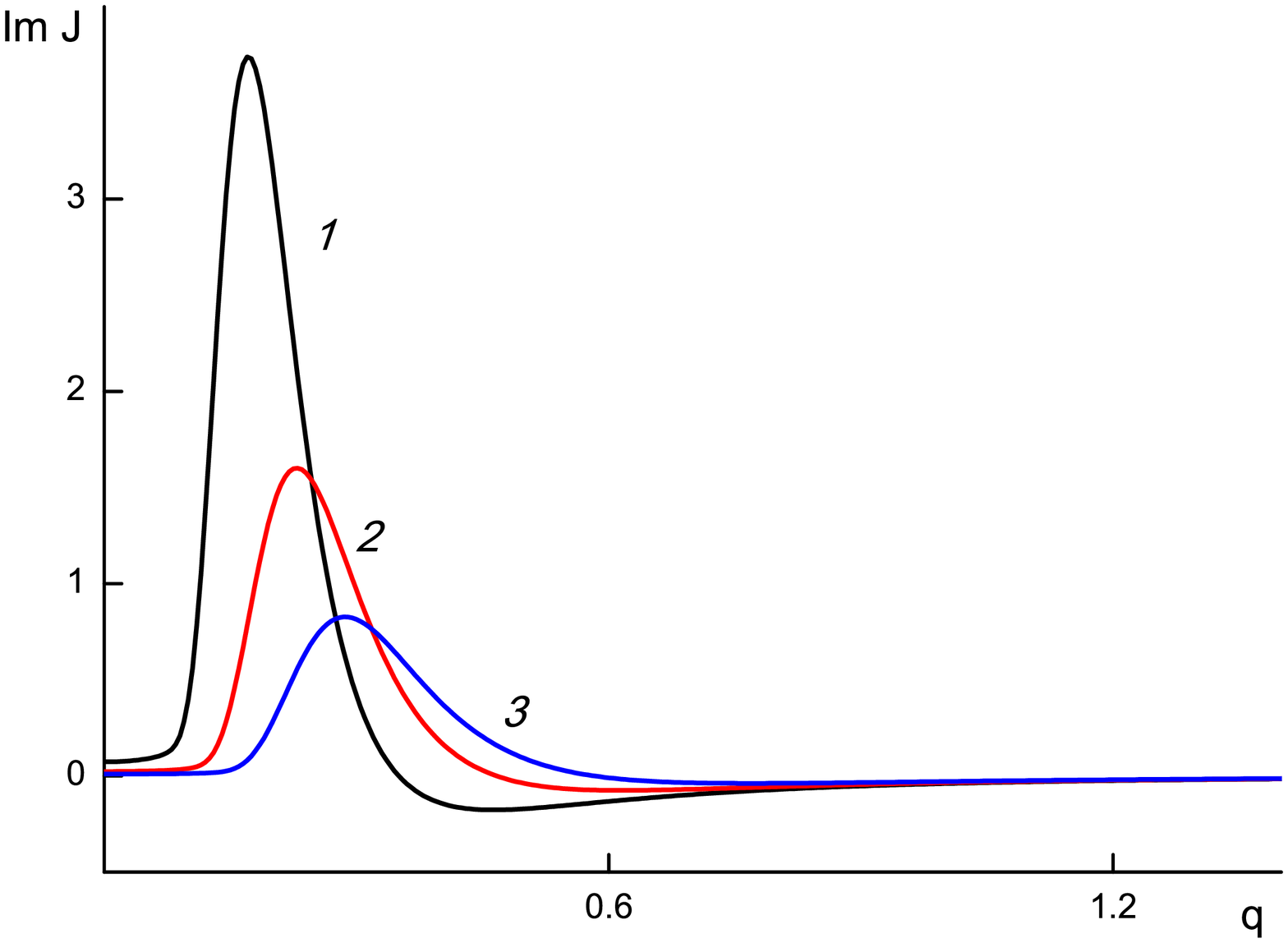}
\center{Fig. 6. Imaginary part of dimensionless density of longitudinal current,
$x=1, \alpha=0$. Curves $1,2,3$ correspond to values of
dimensionless frequency of oscillations of electromagnetic field
$\Omega=0.3, 0.4, 0.5$.}
\end{figure}

\end{document}